\documentclass{article}
\usepackage[final]{neurips_2025}

\makeatletter
\providecommand{\@trackname}{}
\providecommand{\@noticestring}{}
\makeatother
\usepackage{xcolor}
\usepackage{colortbl}
\usepackage{url}
\usepackage{soul}
\usepackage{listings}
\usepackage{graphicx}
\usepackage{subcaption}
\usepackage{caption}
\usepackage{float}
\usepackage{booktabs}
\usepackage{wrapfig}
\usepackage{array}
\usepackage{longtable}
\usepackage{tabularx}

\usepackage{ifpdf}
\usepackage{soul}
\usepackage{amsmath}

\lstdefinestyle{mystyle}{
  language=Ruby,
  basicstyle=\ttfamily\small,
  keywordstyle=\color{blue},
  stringstyle=\color{orange},
  commentstyle=\color{gray},
  breaklines=true,
  frame=single
}

\definecolor{myblue}{rgb}{0.91, 0.96, 0.97}
\definecolor{myyellow}{rgb}{1, 1, 0.93}

\newcommand{\dataset}{27~}

\def\papertitle{Embedding Alignment in Code Generation for Audio}
\def\firstauthor{Sam Kouteili}
\def\secondauthor{Hiren Madhu}
\def\thirdauthor{George Typaldos}
\def\fourthauthor{Mark Santolucito}
\newif\ifpdf
\ifx\pdfoutput\relax
\else
   \ifcase\pdfoutput
      \pdffalse
   \else
      \pdftrue
  \fi
\fi

\ifpdf % compiling with pdflatex
  \usepackage[pdftex,
    pdftitle={\papertitle},
    pdfauthor={\firstauthor, \secondauthor, \thirdauthor, \fourthauthor},
    bookmarksnumbered, % use section numbers with bookmarks
    pdfstartview=XYZ % start with zoom=100% instead of full screen; 
   ]{hyperref}
  \usepackage[figure,table]{hypcap}

\else % compiling with latex
  \usepackage[dvips,
    bookmarksnumbered, % use section numbers with bookmarks
    pdfstartview=XYZ % start with zoom=100% instead of full screen
  ]{hyperref}  % hyperrefs are active in the pdf file after conversion

  \usepackage[dvips]{epsfig,graphicx}
  \graphicspath{{./figures/}}
  \DeclareGraphicsExtensions{.eps}

  \usepackage[figure,table]{hypcap}
\fi

\hypersetup{
    colorlinks,%
    citecolor=black,%
    filecolor=black,%
    linkcolor=black,%
    urlcolor=black
}

\usepackage{multirow}

\title{\papertitle}

\author{
\normalfont
\begin{tabular}{c@{\hspace{1em}}c@{\hspace{1em}}c@{\hspace{1em}}c}
\firstauthor & \secondauthor & \thirdauthor & \fourthauthor \\
Yale University & Yale University & Yale University & Columbia University \\
\scriptsize{\texttt{sam.kouteili@yale.edu}} &
\scriptsize{\texttt{hiren.madhu@yale.edu}} &
\scriptsize{\texttt{george.typaldos@yale.edu}} &
\scriptsize{\texttt{msantolu@barnard.edu}}
\end{tabular}
}

\begin{document}
%
% \capstartfalse
\maketitle
% \capstarttrue
%
\begin{abstract}
Large Language Model (LLM) code generation has the potential to enhance creative coding by allowing users to focus on structural and musical motifs rather than syntactic details. For live-coding and other music-oriented settings, users would benefit from diverse candidates that reflect meaningful differences in the resulting audio. However, current models struggle to produce such diversity, as they lack direct insight into the code’s sonic output and are typically evaluated using text-based similarity metrics. In this paper, we propose a predictive MLP model that learns an embedding alignment map between code and audio, enabling reasoning about musical similarity directly from code embeddings. This alignment introduces musical awareness into code generation workflows, supporting more perceptually relevant candidate selection and opening the door to musically informed code assistants.
\end{abstract}

\section{Introduction}

Creative coding endeavors are emerging as a vibrant space at the intersection of art and computation. One such example is \textit{live-coding}, where performers write music-generating code in real-time. 
% In live-coding performances, code is often displayed as it is written, with insight into the music generation process part of the performance~\cite{toplap2015manifesto}. 
This can often be challenging, as performers need to write syntactically correct code under both time constraints and the pressure of an audience.
% As per the TOPLAP Manifesto, live-coding performers write and revise multimedia-generating code in real time, treating the code as a form of artistic expression~\cite{toplap2015manifesto}.
% Here, the emphasis is on creating expressive and real-time audiovisual performance. 

Recent advances in code generation with LLMs~\cite{codegen1, codegen2, tong_codejudge_2024} present an exciting opportunity for such domains. By lifting much of the syntactic burden, LLMs allow live-coders to focus on higher-level creative motifs and musical ideas. 
However, existing code generation models struggle to provide diverse code candidates in multi-modal domains, where code output is not text, as they do not possess mechanisms to semantically process multi-modal output~\cite{vasilakis2024pretmusic}.
Such models typically evaluate candidate outputs using text-based similarity metrics, which do not capture perceptual or semantic audio differences\footnote{We discuss these techniques more in Appendix Section~\ref{sec:rel_works}.}.

Embedding models offer a potential path forward. Embedding spaces represent meaningful relationships in a given domain by mapping entities to a high-dimensional vector space where ``similar" items are less distant. 
In the context of live-coding, building an alignment map between code and audio would offer a mechanism to reason about musical similarity based only on produced code.

After an initial exploration of the code-audio embedding latent space relationship, we propose a dual Multi-Layer Perceptron (MLP) framework that aligns code and audio embedding spaces. Such a model can provide insights into the topology of the code-audio relationship, helping bridge what an LLM writes and what a user hears. We conduct a study simulating code completion for melody, drum, and bass generation, showing that even on code artifacts with major overlaps, our proposed MLP distinguishes distinct musical semantics only with source code. This highlights our model's potential as a supplement to code completion environments, augmenting them with the ability to reason about code candidates in the auditory domain.

% We then show even with code files with major overlap, the proposed MLP is able to distinguish between distinct music files using only the code. This makes the method more efficient than generating the music and then doing the comparison. Such a model can provide insights into the topology of the code-audio relationship and help bridge what a model writes and what a user hears, potentially employed to supplement code completion/suggestion environments.

\section{Preliminary Investigation}\label{sec:prelim}

\begin{figure}[t]
    \hspace{0.02\columnwidth}
  \begin{subfigure}{.47\columnwidth}
    \centering
    \includegraphics[width=0.7\linewidth]{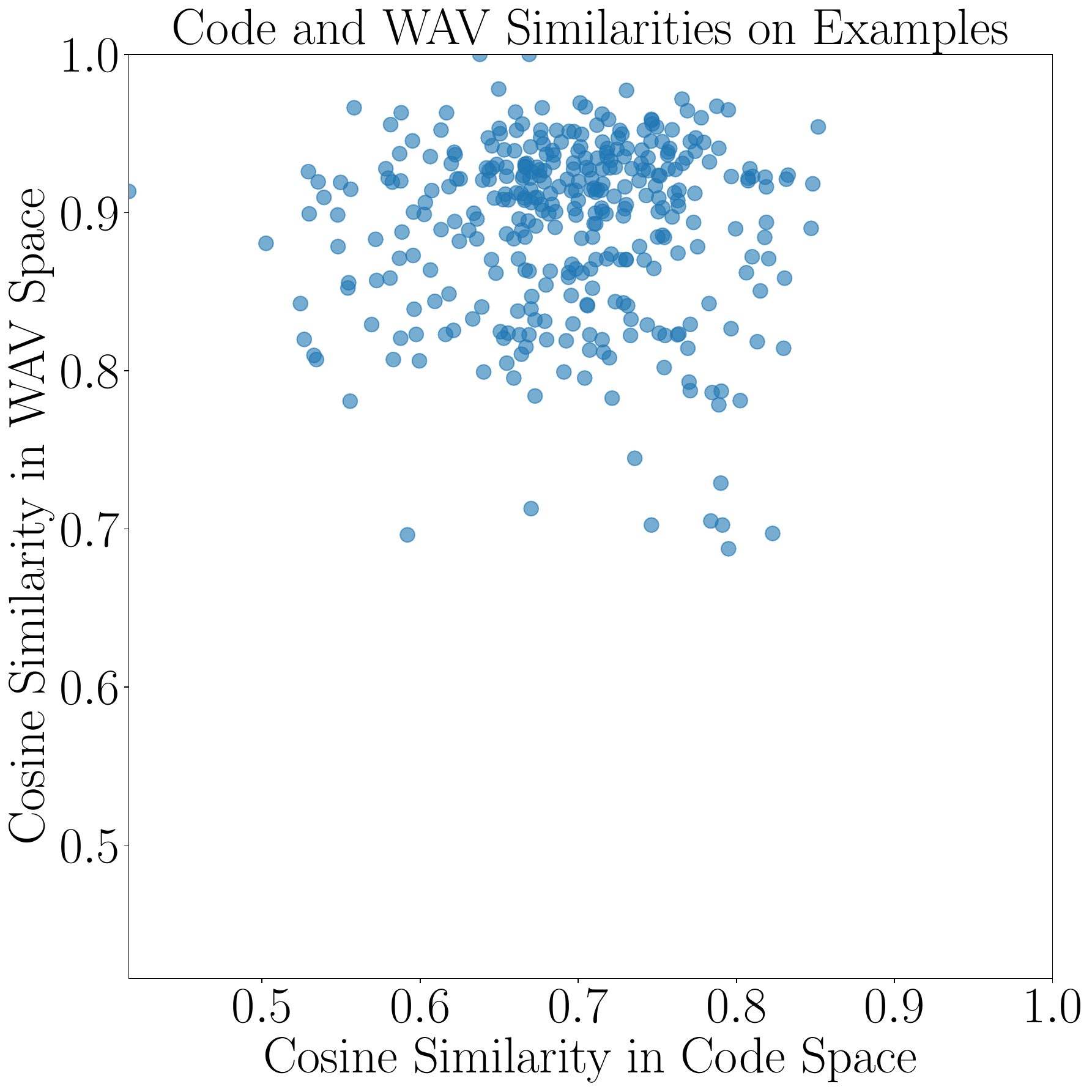}
    \caption{Similarity between Sonic Pi tutorial entries.}
    \label{fig:correlation}
  \end{subfigure}\hspace{-0.02\columnwidth}
  \begin{subfigure}{.47\columnwidth}
    \centering
    \includegraphics[width=0.7\linewidth]{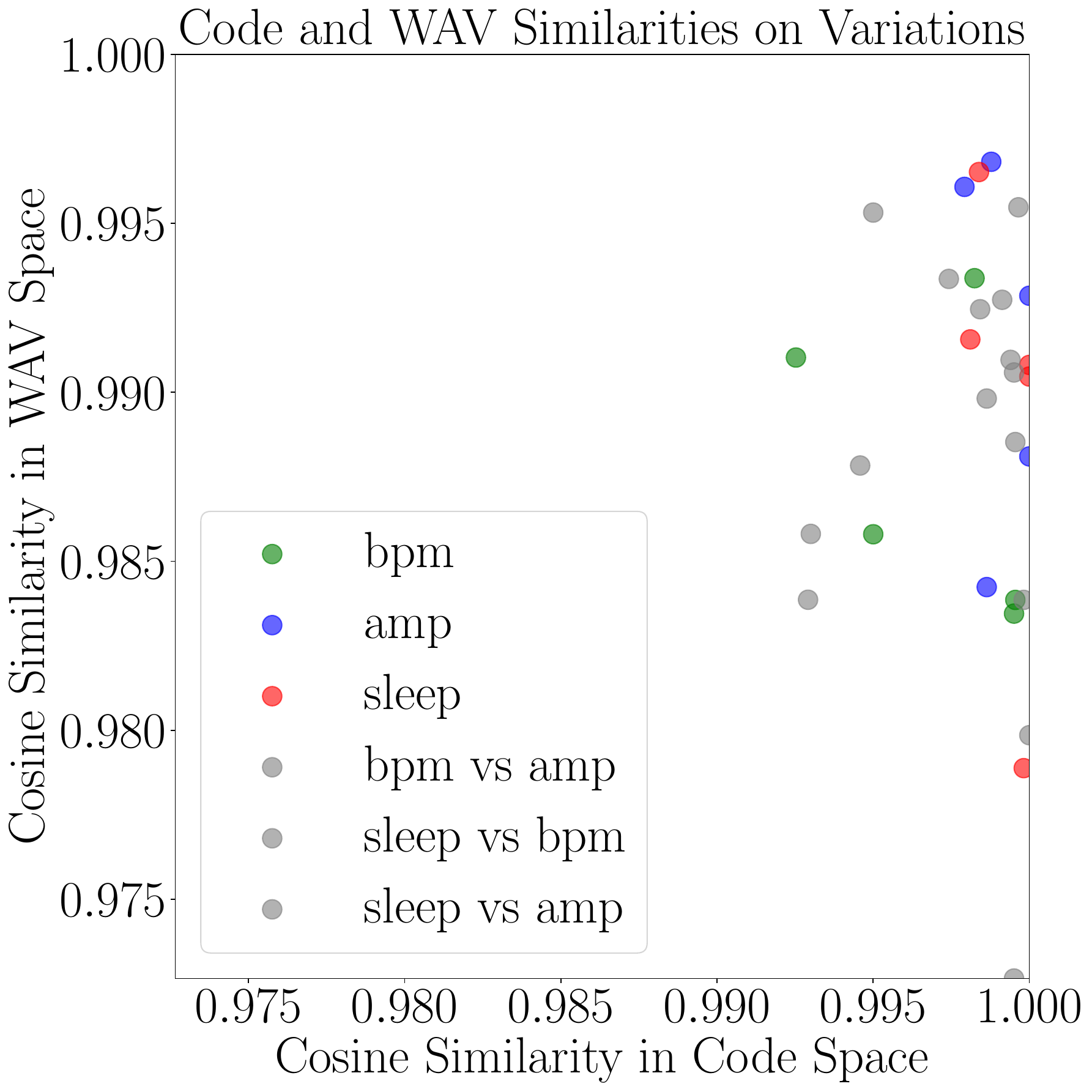}
    \caption{Similarity of small code variations.}
    \label{fig:vari_coor}
  \end{subfigure}
  \caption{Distances between sample embeddings show nontrivial alignment mapping. We extract code and audio embeddings with \texttt{distilroberta-base} and  \texttt{wav2vec2} respectively. Audio is clipped to 9 measures at 120 BPM. Embedding similarity is computed as vector cosine similarity.}
  \label{fig:side_by_side}
\end{figure}

As an initial exploration, we investigate the latent relationship between code and audio embedding spaces.
% hoping to gain a better sense of our latent space. 
For our first study, we select the \dataset Sonic Pi tutorial entries and record \dataset corresponding audio outputs, collecting code and audio embeddings.  
We chose Sonic Pi due to its prominence in the live-coding community, terseness, and strong documentation. For each entry (code+audio) in the dataset, we plot code and audio embedding distances to all other entries, comparing code to code and audio to audio. 
%, with a preliminary investigation finding that code-generation models seem to perform better with it than other computer-music DSLs.
% The dataset we consider consists of \dataset distinct Sonic Pi programs and \dataset corresponding audio files. 
% For each audio clip, we extract the first nine measures at a fixed tempo of 120 BPM.
% ensuring that audio content is captured in a standardized manner. 
% Using an automated script, we simultaneously load both the audio clips and their associated program code, computing embeddings for each entry.
%We extract code and audio embeddings with \texttt{distilroberta-base}~\cite{DistilBERTAD} and  \texttt{wav2vec2}~\cite{wav2vec} respectively. We assess embedding similarity by considering vector cosine similarity. 
% We compute Pearson and Spearman correlation coefficients based on the similarity scores of the samples.
Figure \ref{fig:correlation} displays our results: looking at our findings, no evident relationship emerges. 
Low Pearson (0.0159, p=0.6770) and Spearman correlations (0.0409, p=0.4450) affirm no linear or rank-order relationship between the embedding spaces. 
% These results match our conceptual intuition of this embedding map: a perfect alignment would fully encode the Sonic Pi compiler and audio engine.

% \begin{figure}[htbp]
%     \centering
%     \begin{subfigure}[c]{0.4\textwidth}
%         \centering
%         \includegraphics[width=0.3\textwidth]{figures/similarity_correlation.pdf}
%         \caption{Embedding distances between pairs of SonicPi Examples.}
%         \label{fig:correlation}
%     \end{subfigure}
%     \hfill
%     \begin{subfigure}[c]{0.4\textwidth}
%         \centering
%         \includegraphics[width=0.3\textwidth]{figures/variations_correlation_all.pdf}
%         \caption{Embedding distances between all pairs of small code variations.}
%         \label{fig:vari_coor}
%     \end{subfigure}
%     \caption{Plotting distances between samples in both embedding spaces shows that mapping between spaces is nontrivial.}
%     \label{fig:side_by_side}
% \end{figure}

% preamble: \usepackage{subcaption} (no 'subfig' or 'caption')

% \begin{figure}[htbp]
%     \centering
%     \begin{subfigure}[c]{0.45\textwidth}
%         \centering
%         \includegraphics[width=\linewidth]{figures/similarity_correlation.pdf}
%         \caption{Embedding distances between pairs of SonicPi Examples.}
%         \label{fig:correlation}
%     \end{subfigure}
%     \hfill
%     \begin{subfigure}[c]{0.45\textwidth}
%         \centering
%         \includegraphics[width=\linewidth]{figures/variations_correlation_all.pdf}
%         \caption{Embedding distances between all pairs of small code variations.}
%         \label{fig:vari_coor}
%     \end{subfigure}
%     \caption{Plotting distances between samples in both embedding spaces shows that mapping between spaces is nontrivial.}
%     \label{fig:side_by_side}
% \end{figure}
% With our first exploration yielding no evident mapping, we conduct a more fine-grained study investigating the sensitivity of code and audio embeddings to small program code modifications. 
We proceed to investigate the sensitivity of code and audio embeddings to small program code modifications. We modify the six largest programs in the tutorials by varying the values of three parameters: sleep time, amplitude (amp), and beats per minute (bpm). 
We posit that minor code modifications should yield similar code embeddings, but may result in varying audio embeddings depending on the altered variable. 
Changes in amplitude, for example, should affect audio embeddings less than changes in sleep, which may change stem syncopation.
% Figure \ref{fig:vari_coor} displays the cosine similarities of minor code variations in the code and audio embedding space. 
Figure \ref{fig:vari_coor} confirms that fuzzed code artifacts maintain a high ($>0.990$) code embedding similarity. 
While similarity is also high in the audio embedding space, the range is larger, with similarity scores below 0.975. 
Interestingly, there is no perceptible trend between the modified variable and the resultant audio embedding: sleep, bpm, and amp variations exhibit inconsistent changes in the audio embedding space.
%As was the case with the previous study, we have low Pearson and Spearman correlations. 
The associative, albeit not equivalent, range compression of code and audio embedding distances suggests some coarse association between the domains; however, such an association is not trivial.

% \sam{need to compute pearson and spearman for fine grained. also what were the ranges considered for bpm, sleep, amp?}      
% \input{sections/3-data}
\section{Model Implementation}
% To train an embedding alignment model, we require a large amount of Sonic Pi code and corresponding audio samples. 
% The best candidate for a Sonic Pi dataset comes from the \dataset Sonic Pi tutorial entries, written or selected by the language's creator, Sam Aaron. 
% While the dataset is clean and diverse, it is not sufficiently populated.
% We thus explore dataset augmentation via templates. 
% Using the Jinja engine \cite{jinja}, we transform each of the entries into a template and randomize various parameters. 
% A template example can be seen in \autoref{lst:compus-temp} with a list of the parameters used in \autoref{tab:temp-params}.
% We render 500 different Sonic Pi code files for each one of the \dataset examples, generating 13,500 entries. 

To train an embedding alignment model, we utilize the \dataset Sonic Pi code entries mentioned in \autoref{sec:prelim}, along with the Jinja template engine \cite{jinja} to augment the dataset and randomize various parameters. A table of parameters used for templating is provided in \autoref{tab:temp-params}, and a template example can be seen in \autoref{lst:compus-temp}.
We render 500 different Sonic Pi code files, generating a total of 13,500 entries. 
We use \texttt{distilroberta-base} \cite{DistilBERTAD} to generate code embeddings and Meta's \texttt{wav2vec2} \cite{wav2vec} for the respective audio embeddings.
%Audio embeddings were consistently sampled at a 16 kHz rate.
% We create an arrow dataset which contains the code \textit{.pi} descriptions, the code embeddings, and the audio embeddings of 5400 samples, filtering faulty entries. 
% This dataset is available \href{https://github.com/anonymizedforsubmission}{here[anonymized]}
% To align code and audio embeddings into a shared latent space, 
We adopt a symmetric architecture consisting of two independent MLPs: one for code embeddings $\texttt{MLP}_c$ and one for audio embeddings $\texttt{MLP}_a$. 
Both networks take as input the respective modality's pretrained embeddings and project them into a common embedding space of dimension $d_{\text{out}}$.

Each MLP consists of $L$ linear layers, with intermediate hidden layers of dimension $d_{\text{hidden}}$, each followed by \texttt{BatchNorm} and \texttt{GELU} activations. We use \texttt{BatchNorm} to stabilize training by normalizing activations across the batch, and \texttt{GELU} as the activation function due to its smooth, non-linear behavior that improves gradient flow and empirical performance over ReLU in deep networks. We project the pre-trained code and audio embeddings as 
$c_i = \texttt{MLP}_c(c^0_i), a_i = \texttt{MLP}_a(a^0_i)$
where, $c^0_i$ and $a^0_i$ are the embeddings extracted from the pre-trained model, and $c_i$ and $a_i$ are the aligned embeddings.
This formulation was selected to capture non-linear transformations without introducing architectural biases toward either modality. 
Unlike attention-based architectures, MLPs efficiently map pre-trained embeddings into an aligned representation space.

To train the models, we employ \textbf{InfoNCE} loss, a contrastive learning objective that brings semantically aligned code-audio pairs closer while pushing apart mismatched pairs in the same batch. This choice is motivated by the need for self-supervised alignment, where explicit labels are not available, but semantic consistency can be inferred from pairing.
Given a batch of $N$ aligned code-audio embeddings $\{(c_i, a_i)\}_{i=1}^N$, cosine similarity is defined as:
\[
\scalebox{0.9}{$
\text{sim}(c_i, a_j) = \frac{c_i^\top a_j}{\|c_i\| \cdot \|a_j\|}.
$}
\]
% shown in \autoref{eq:loss}.
% \begin{equation}
% \text{sim}(c_i, a_j) = \frac{c_i^\top a_j}{\|c_i\| \cdot \|a_j\|}.
% \end{equation}
The InfoNCE loss for a single positive pair $(c_i, a_i)$ and the overall loss are computed as shown in \autoref{eq:loss}, with $\tau$ being the temperature hyperparameter that controls the sharpness of the similarity distribution.
This contrastive formulation applies well to this setting, where each code-audio pair is semantically meaningful but hard supervision is unavailable. InfoNCE encourages the model to preserve pairwise relationships and learn embeddings that are useful for downstream retrieval and matching tasks.
% \begin{equation}
% \mathcal{L}_i = -\log \frac{\exp(\text{sim}(c_i, a_i)/\tau)}{\sum_{j=1}^N \exp(\text{sim}(c_i, a_j)/\tau)}, \quad \mathcal{L} = \frac{1}{N} \sum_{i=1}^N \mathcal{L}_i,
% \end{equation}
%where $\tau$ is a temperature hyperparameter that controls the sharpness of the similarity distribution. 
\begin{equation}\label{eq:loss}
\scalebox{0.9}{$
\mathcal{L}_i = -\log \frac{\exp(\text{sim}(c_i, a_i)/\tau)}{\sum_{j=1}^N \exp(\text{sim}(c_i, a_j)/\tau)}, 
\qquad
\mathcal{L} = \frac{1}{N} \sum_{i=1}^N \mathcal{L}_i
$}
\end{equation}

\section{Experiments}

% To quantify the alignment between learned representations, we utilize two additional similarity metrics: \textbf{Canonical Correlation Analysis (CCA)} and \textbf{Centered Kernel Alignment (CKA)}.
% \textit{CCA} measures the maximum linear correlation between two multivariate random variables after projecting them onto a shared subspace.  
% In our context, given code and audio embeddings $C$ and $A$, CCA finds linear projections such that the correlation between $Cw_c$ and $Aw_a$ is maximized.  
% The resulting correlation scores reflect the extent to which a linear transformation can align the two modalities.  
% To ensure comparability across configurations, we normalize CCA scores by the embedding dimensionality.
% \textit{CKA}, on the other hand, captures similarities between representations in a way that is invariant to orthogonal transformations and isotropic scaling.

\noindent\textbf{Setup:} To quantify the alignment between learned representations, we utilize two similarity metrics.
First, Canonical Correlation Analysis (CCA) measures the maximum linear correlation between two multivariate random variables after projecting them onto a shared subspace.
Given code and audio embeddings $C$ and $A$, CCA finds linear projections such that the correlation between $Cw_c$ and $Aw_a$ is maximized and the resulting correlation scores reflect the extent to which a linear transformation can align the two modalities.
Second, Centered Kernel Alignment (CKA) captures similarities between representations in a way that is invariant to orthogonal transformations and isotropic scaling.
Unlike CCA, which measures linear alignment, CKA is sensitive to more general (nonlinear) structural similarities. We explain CKA in more detailore detail in \autoref{sec:CKA}.
By combining InfoNCE-based contrastive training with post-hoc evaluation using CKA and normalized CCA, we comprehensively assess the degree to which learned embeddings from the code and audio modalities are aligned at both linear and structural levels. We run 24 configurations varying hidden/output dimension, layers, and learning rate. Metrics are averaged over five runs. 

We selected the best model and evaluated alignment quality using neighborhood-based metrics such as Jaccard similarity and top-$k$ overlap. These measures assess whether nearest neighbors in the code embedding space correspond to nearest neighbors in the audio embedding space. We explain the data generation for this evaluation in \autoref{tab:experiment-setup}.

\noindent\textbf{Results:} \autoref{tab:eval} reports results of hyperparameter tuning for aligning code and waveform embeddings using a contrastive framework. The first row shows pre-alignment baselines, with low CKA (0.090) and CCA (0.140), indicating minimal correlation between raw embeddings.
Post-alignment, the best configuration achieved a CKA of 0.590 (Config. 21) and a normalized CCA of 0.902 (Config. 24), representing over six-fold improvements in both metrics. These gains demonstrate that the model learns a meaningful shared embedding space, enabling reliable approximation of audio embeddings from code even without explicit supervision.

We evaluate our model against a raw code-based baseline across three scenarios,\textbf{melody}, \textbf{drum}, and \textbf{bass}, where we simulate LLM-assisted code completion for melody, drum pattern, and bassline addition.
%, using cosine similarity (Euclidean in one case). 
(see Table~\ref{tab:experiment-setup} for experimental setup). We report Jaccard, overlap@3, and rank correlations (Spearman, Pearson). 
\autoref{tab:results} summarizes our results.

\begin{table}[h]
\centering
\begin{tabular}{lcccccc}
\toprule
\textbf{Scenario} & \textbf{Method} & \textbf{Jaccard} & \textbf{Overlap@3} & \textbf{Spearman} & \textbf{Pearson} \\
\midrule
\multirow{2}{*}{melody} 
 & Raw   & 0.20 & 0.33 & \textbf{0.21} & \textbf{0.18} \\
 & Ours  & \textbf{0.34} (0.21) & \textbf{0.47} (0.27) & 0.16 (0.17) & 0.07 (0.18) \\
\midrule
\multirow{2}{*}{drum} 
 & Raw   & 0.00 & 0.00 & \textbf{	} & -0.25 \\
 & Ours  & \textbf{0.16} (0.08) & \textbf{0.27} (0.13) & -0.05 (0.18) & \textbf{-0.12} (0.20) \\
\midrule
\multirow{2}{*}{bass} 
 & Raw   & 0.20 & 0.33 & 0.24 & 0.21 \\
 & Ours  & \textbf{0.50} (0.00) & \textbf{0.67} (0.00) & \textbf{0.44} (0.05) & \textbf{0.46} (0.05) \\
\bottomrule
\end{tabular}
\caption{Comparison of raw baseline vs. our method across three scenarios. Standard deviations for our method are reported in parentheses.}
\label{tab:results}
\end{table}

\noindent\textbf{Findings:}  
Across all three settings, our method consistently improves neighborhood-based metrics (Jaccard and overlap@3), with the largest gains on \textbf{s2-drum} (where the baseline fails entirely) and \textbf{s3-bass} (where both selection and correlation improve markedly). Even in \textbf{s1-mel}, where fine-grained correlations dip slightly, top-$k$ accuracy improves, highlighting complementary strengths of the evaluation metrics. Importantly, these gains are achieved \textit{directly from code embeddings}, without compiling audio or extracting audio embeddings, a process that is computationally expensive and time-consuming. This demonstrates the efficiency and practicality of our approach.

\begin{figure}[H]
    \centering
    \begin{subfigure}[b]{0.48\textwidth}
        \centering
        \begin{subfigure}[b]{0.48\textwidth}
            \includegraphics[width=\textwidth]{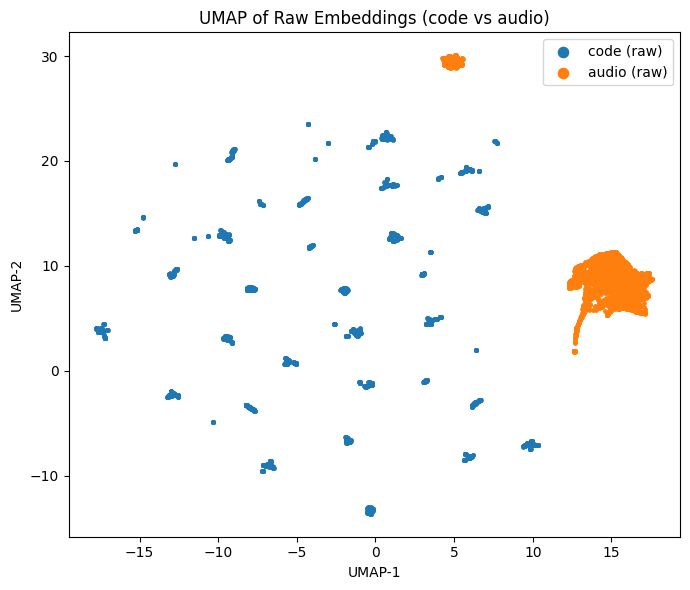}
            % \caption{Caption for image 1}
            \label{fig:umap-raw}
        \end{subfigure}
        \hfill
        \begin{subfigure}[b]{0.48\textwidth}
            \includegraphics[width=\textwidth]{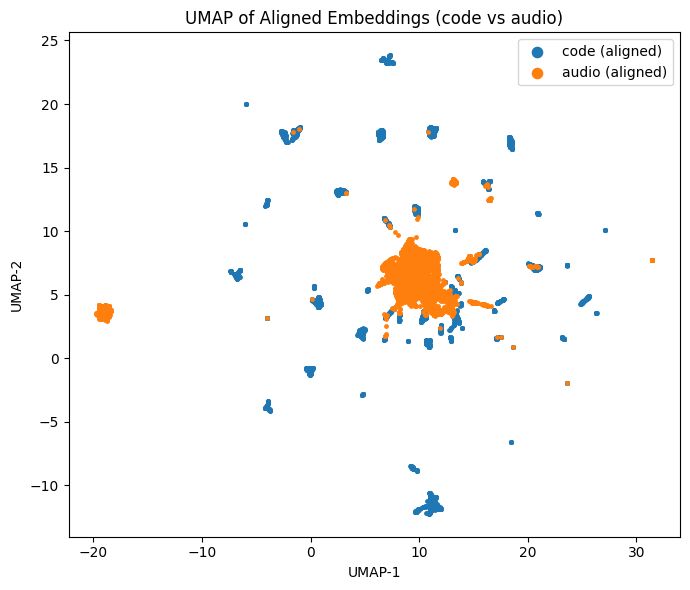}
            % \caption{Caption for image 2}
            \label{fig:umap-aligned}
        \end{subfigure}
        \vspace{-0.05\linewidth}
        \caption{UMAP visualization of code and audio embeddings before and after alignment.}
        \label{fig:umap}
    \end{subfigure}
    \hfill
    \begin{subfigure}[b]{0.48\textwidth}
        \centering
        \begin{subfigure}[b]{0.48\textwidth}
            \includegraphics[width=\textwidth]{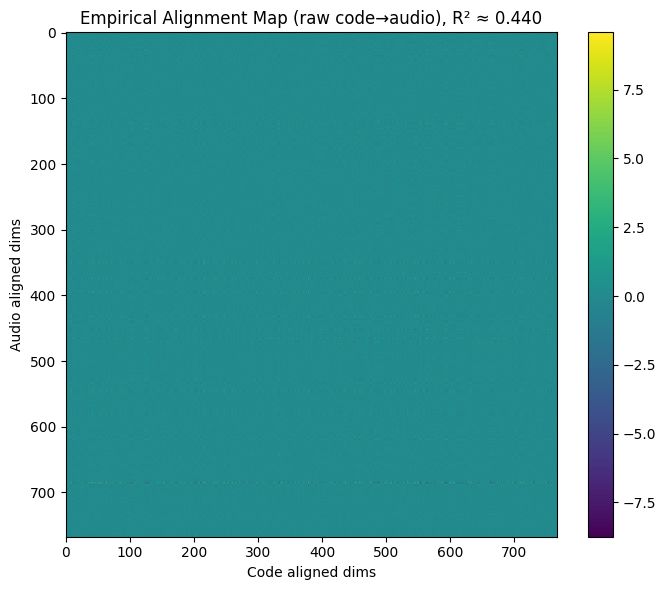}
            % \caption{Caption for image 3}
            \label{fig:heatmap-raw}
        \end{subfigure}
        \hfill
        \begin{subfigure}[b]{0.48\textwidth}
            \includegraphics[width=\textwidth]{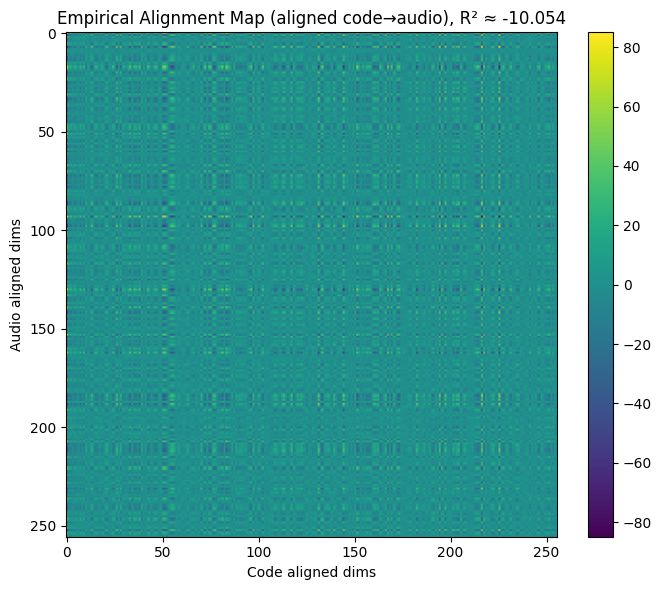}
            % \caption{Caption for image 4}
            \label{fig:heatmap-aligned}
        \end{subfigure}
        \vspace{-0.05\linewidth}
        \caption{Empirical alignment heat maps showing code-audio correspondence quality}
        \label{fig:heatmap}
    \end{subfigure}
    \caption{UMAP and Empirical heat map visualizations demonstrate improved clustering and code-audio matching after alignment training.}
    \label{fig:results}
\end{figure}

In Figure~\ref{fig:results}, we illustrate the effectiveness of our model in bridging the semantic gap between code and audio modalities. The UMAP visualizations in Figure~\ref{fig:umap} reveal that raw embeddings exhibit complete modal separation, with code (blue) and audio (orange) occupying entirely distinct regions of the embedding space. After alignment, we observe overlap between modalities, with audio embeddings clustering within code neighborhoods, demonstrating successful semantic bridging. 

Interestingly, the empirical alignment maps in Figure~\ref{fig:heatmap} show that while raw embeddings achieve higher linear correlation ($R^2 = -10.054$ to $0.840$), this linear fit fails to capture true cross-modal semantics, as evidenced by the poor clustering in UMAP space. Our aligned model sacrifices some linear correlation for semantically meaningful overlap, where related code-audio pairs now occupy shared embedding regions. This nonlinear alignment approach successfully maps semantically related content across modalities into proximate embedding neighborhoods, enabling effective cross-modal retrieval and generation despite reduced linear correlation measures.

%This demonstrates that traditional linear metrics inadequately assess cross-modal alignment quality, 

%—our nonlinear alignment approach successfully maps semantically related content across modalities into proximate embedding neighborhoods, enabling effective cross-modal retrieval and generation despite reduced linear correlation measures.

% \begin{figure}[h]
%     \centering
%     \begin{subfigure}[b]{0.24\textwidth}
%         \includegraphics[width=\textwidth]{figures/umap-raw.jpeg}
%         \caption{Caption for image 1}
%         \label{fig:image1}
%     \end{subfigure}
%     \hfill
%     \begin{subfigure}[b]{0.24\textwidth}
%         \includegraphics[width=\textwidth]{figures/umap-aligned.jpeg}
%         \caption{Caption for image 2}
%         \label{fig:image2}
%     \end{subfigure}
%     \hfill
%     \begin{subfigure}[b]{0.24\textwidth}
%         \includegraphics[width=\textwidth]{figures/empirical-raw.jpeg}
%         \caption{Caption for image 3}
%         \label{fig:image3}
%     \end{subfigure}
%     \hfill
%     \begin{subfigure}[b]{0.24\textwidth}
%         \includegraphics[width=\textwidth]{figures/empirical-aligned.jpeg}
%         \caption{Caption for image 4}
%         \label{fig:image4}
%     \end{subfigure}
%     \caption{Overall caption for all four images}
%     \label{fig:four_images}
% \end{figure}
% \clearpage
\section{Conclusion}

In this work, we construct a code–audio embedding alignment map to bridge the cross-modal semantic gap for code generation models (LLMs).
Our preliminary analysis reveals minimal linear or rank-order relationships between the respective latent spaces, thereby motivating the development of a non-linear alignment model.
Leveraging Sonic Pi templates, we augment a curated dataset of code–audio embeddings and train a dual MLP model architecture to project these embeddings into a shared latent space. Model evaluation and hyperparameter tuning are quantified using CCA and CKA.
Our final experiments, conducted on three distinct live-coding code assistance tasks, demonstrate that the proposed model effectively captures and distinguishes auditory differences among code candidates.

This study presents an initial step toward an integrated framework designed to support live coders with code generation models.
Future work will focus on extending this framework by incorporating our alignment map into a generative code assistant environment to better realize artistic intent.

% In this work, we examine the ability of LLMs to assist users with live coding.
% We aimed to show whether a model can produce variable code suggestions based on code and audio embeddings.
% To investigate this hypothesis, we utilized the Sonic Pi live coding framework and created a curated dataset of code-audio embeddings,  using templates. 
% We used two independent Multi-Layer Perceptrons and quantified the evaluation results using the Canonical Correlation Analysis and Centered Kernel Alignment metrics.
% Our results showcase that our neural network model successfully learns to align code and audio embeddings and can reliably approximate the audio embedding given the corresponding code embedding.

% \sam{toadd: this is an initial work building towards an integrated framework that seeks to support live-coders with code generation models. Future works would explore constructing this framework and integrating this alignment map with a generative }
% \clearpage

% \printbibliography{}

%%%%%%%%%%%%%%%%%%%%%%%%%%%%%%%%%%%%%%%%%%%%%%%%%%%%%%%%%%%%%%%%%%%%%%%%%%%%%
%bibliography here
\newpage
\bibliographystyle{plainnat}
\bibliography{sources}
\newpage
\renewcommand{\thefigure}{A\arabic{figure}}
\renewcommand*{\thepage}{A\arabic{page}}
\renewcommand{\thetable}{A\arabic{table}}
\setcounter{figure}{0}
\setcounter{table}{0}
\setcounter{page}{1}
\appendix
\section{Related Work}\label{sec:rel_works}

\subsection{Code Generation Models}

With the rise of large language models (LLMs), systems such as GitHub’s Copilot~\cite{copilot} and OpenAI’s Codex~\cite{Chen2021Codex} have demonstrated strong code generation capabilities for complex programming tasks, having been trained on extensive code repositories and fine-tuned to capture code semantics.
Code generation models are typically evaluated on competitive programming tasks, with the HumanEval dataset~\cite{Chen2021Codex} serving as one of the predominant benchmarks.
Recent research has expanded these efforts to domains such as music computing and the study of coding-related sentiments~\cite{wang_2024}.
However, given the inherent non-determinism of LLMs, ensuring code correctness and consistency remains a persistent challenge~\cite{OuYang2024NonDeterminism}.
These challenges are further exacerbated in creative coding contexts, where limited training data exist for certain domain-specific languages (DSLs).
Moreover, LLMs have been shown to struggle with music comprehension~\cite{vasilakis2024musicEvalLLM}, suggesting corresponding difficulties in evaluating code for computer music generation.

\subsection{Music Programming Languages}
%TODO: rewrite

Many modern music programming languages take the form of DSLs (domain specific languages) - languages that are designed for a specific application domain. 
One such example is SuperCollider~\cite{SuperCollider}, an audio programming language and environment for real-time audio synthesis and algorithmic composition. 
Conversely, FAUST (Functional AUdio STream)~\cite{orlarey:hal-02159014} is a purely functional programming language for real-time signal processing. 
Live coding languages are a subset of music programming languages tailored for live music performance. 
One example is Sonic Pi~\cite{Aaron2024SonicPi}, a live coding language built on Ruby that has been prominently adopted by the community. Sonic Pi has shown promise in educational settings, introducing students to computer science concepts through real-time music coding. 
Tidal Cycles~\cite{tidal} is another functional alternative built on the Haskell functional programming language; it offers programmers a declarative approach to live coding. 
Strudel~\cite{strudel} is a variant of Tidal Cycles built on JavaScript garnering notable community adoption.
% As is the case with general purpose music programming languages, live coding languages make design decisions with the aim of providing a better interface to users to express their intention.
% Our work is similar to language design in that we explore the potential of LLMs to make live coding less syntactically burdened.

\subsection{Embedding Space Alignment}

Embedding models attempt to produce learned dimensional representations of data in a hyperplane. 
Pre-trained embedding models map data - text, images, audio, programs - into vector spaces that encode relational semantics. 
Program embedding models have been applied to augment code-classification and auto-completion~\cite{10691715}. 
Similarly, audio embedding models have been considered for speech recognition, music generation, and audio classification~\cite{Ozkaya2020audioCaptioning}. Audio embedding models capture acoustic features and temporal dependencies, with different models highlighting different auditory features~\cite{baevski2020wav2vec, cramer2019openl3}. 
%The nature of embeddings makes it so that encoded variations may not reflect perceptible differences - certain works have investigated this fact in both audio and program domains~\cite{Elizadle2019CrossModal}.

Given two distinct embedding latent spaces, one may inquire about an alignment mapping relationship. Early alignment methods applied to language and knowledge graphs analytically established mappings between these spaces; however, recent efforts on more complex maps have employed unsupervised methods~\cite{Biswas2020EmbeddingAlignment}. With the rise of multi-modal LLMs, cross-modal alignment has begun to appear in audiovisual domains~\cite{Elizadle2019CrossModal}. Novel works have explored ways of encoding linguistic semantics in audio embeddings, and have even presented joint embedding spaces between the two fields~\cite{devnani2024learningspatiallyawarelanguageaudio, huang2022mulanjointembeddingmusic}.

\newpage

\section{Dataset Augmentation}

\subsection{Templating Parameters}

\begin{table}[h]
    \centering
    \caption{Parameters used for templating}
    \begin{tabular}{cc}
        \toprule
        \textbf{Parameters} & \textbf{Example Values} \\
        \midrule
        samples & ambi\_choir, bd\_haus \\
        synths & beep, rodeo \\
        character & major, minor \\
        attack/release\_range & [0, 10] \\
        amp\_range & [0, 10] \\
        sleep\_range & [0.1, 5.0] \\
        effects & echo, compressor \\
        notes & C2, Db2, $\ldots$, C6 \\
        \bottomrule
    \end{tabular}
    \label{tab:temp-params}
\end{table}

\subsection{Sample Templated Dataset Entry} % unnumbered title

% % If you want it to appear in the TOC:
% \addcontentsline{toc}{section}{Appendix}

% \begin{figure}[ht]
%     \centering
%     \lstinputlisting[
%         style=mystyle,
%         caption={Jinja Template used for the Compus Beats example},
%         label={fig:compus-temp}
%     ]{data_temp/compus_beats.j2}
%     \caption{Appendix Entry 1}
% \end{figure}

\begin{lstlisting}[
  style=mystyle,
  caption={Jinja Template used for parametrising Compus Beats dataset entry},
  label={lst:compus-temp},
  float,
  floatplacement=htbp
]
# Compus Beats

# Coded by Sam Aaron

use_sample_bpm :{{samples_bpm[0]}}, num_beats: {{repeat_small_ints[0]}}

live_loop :loopr do
  sample :{{samples_bpm[0]}}, rate: [0.5, 1, 1, 1, 1, 2].choose unless one_in(10)
  sleep {{sleep_values[0]}}
end

live_loop :bass do
  sample :{{sample_values[0]}}, amp: rrand(0.1, 0.2), rate: [0.5, 0.5, 1, 1,2,4].choose if one_in(4)
  use_synth :{{synth_values[0]}}
  use_synth_defaults mod_invert_wave: 1
  play :{{note_values[0]}}, mod_range: 12, amp: rrand(0.5, 1), mod_phase: [0.25, 0.5, 1].choose, release: {{release_values[(1)% release_values|length]}}, cutoff: rrand(50, 90)
  play :{{note_values[(1) % note_values|length]}}, mod_range: [24, 36, 34].choose, amp: {{amp_values[0]}}, mod_phase: 0.25, release: {{release_values[0]}}, cutoff: {{repeat_large_ints[0]}}, pulse_width: rand
  sleep {{sleep_values[(1)% sleep_values|length]}}
end

 
\end{lstlisting}

\clearpage

\section{Model Evaluation}

\subsection{CKA}\label{sec:CKA}
CKA operates on kernel matrices $K$ and $L$ derived from embeddings, as shown in \autoref{eq:cka}, where $K_c$ and $L_c$ are centered versions of the kernel matrices, and $\langle \cdot, \cdot \rangle_F$ denotes the Frobenius inner product.  
A CKA score close to 1 indicates high structural similarity between representations.
\begin{equation}\label{eq:cka}
% \scalebox{0.9}{$
\text{CKA}(K, L) = \frac{\langle K_c, L_c \rangle_F}{\|K_c\|_F \cdot \|L_c\|_F},
% $}
\end{equation}

\subsection{Hyperparameter Tuning}

\begin{table}[h]
\centering
\caption{CCA and CKA comparison across hyperparameters. First row shows pre-alignment metrics. Best post-alignment results are in \textbf{bold} (1st) and \underline{underlined} (2nd).}
\label{tab:eval}
% \resizebox{0.66\columnwidth}{!}{%
\begin{tabular}{ccccccc}
\toprule
\textbf{Config} & \textbf{$d_{hidden}$} & \textbf{$d_{out}$} & \textbf{L} & \textbf{LR} & \textbf{CKA} & \textbf{CCA} \\
\midrule
-- & -- & -- & -- & Before training & 0.090 $\pm$ 0.001 & 0.145 $\pm$ 0.003 \\
1  & 256 & 128 & 5 & 1e-4 & 0.420 $\pm$ 0.019 & 0.523 $\pm$ 0.021 \\
2  & 128 & 64  & 1 & 1e-3 & 0.455 $\pm$ 0.004 & 0.480 $\pm$ 0.004 \\
3  & 128 & 64  & 1 & 1e-4 & 0.463 $\pm$ 0.011 & 0.378 $\pm$ 0.011 \\
4  & 128 & 64  & 3 & 1e-3 & 0.424 $\pm$ 0.024 & 0.510 $\pm$ 0.019 \\
5  & 128 & 64  & 3 & 1e-4 & 0.398 $\pm$ 0.021 & 0.372 $\pm$ 0.010 \\
6  & 128 & 64  & 5 & 1e-3 & 0.422 $\pm$ 0.014 & 0.552 $\pm$ 0.009 \\
7  & 128 & 64  & 5 & 1e-4 & 0.357 $\pm$ 0.040 & 0.396 $\pm$ 0.011 \\
8  & 128 & 128 & 1 & 1e-3 & 0.472 $\pm$ 0.057 & 0.660 $\pm$ 0.021 \\
9  & 128 & 128 & 1 & 1e-4 & 0.490 $\pm$ 0.033 & 0.522 $\pm$ 0.010 \\
10 & 128 & 128 & 3 & 1e-3 & 0.494 $\pm$ 0.017 & 0.691 $\pm$ 0.010 \\
11 & 128 & 128 & 3 & 1e-4 & 0.459 $\pm$ 0.031 & 0.547 $\pm$ 0.003 \\
12 & 128 & 128 & 5 & 1e-3 & 0.410 $\pm$ 0.023 & 0.735 $\pm$ 0.013 \\
13 & 128 & 128 & 5 & 1e-4 & 0.407 $\pm$ 0.055 & 0.571 $\pm$ 0.013 \\
14 & 256 & 64  & 1 & 1e-3 & 0.468 $\pm$ 0.063 & 0.499 $\pm$ 0.011 \\
15 & 256 & 64  & 1 & 1e-4 & 0.514 $\pm$ 0.014 & 0.357 $\pm$ 0.005 \\
16 & 256 & 64  & 3 & 1e-3 & 0.461 $\pm$ 0.035 & 0.556 $\pm$ 0.007 \\
17 & 256 & 64  & 3 & 1e-4 & 0.454 $\pm$ 0.027 & 0.366 $\pm$ 0.006 \\
18 & 256 & 64  & 5 & 1e-3 & 0.432 $\pm$ 0.005 & 0.644 $\pm$ 0.013 \\
19 & 256 & 64  & 5 & 1e-4 & 0.386 $\pm$ 0.014 & 0.372 $\pm$ 0.010 \\
20 & 256 & 128 & 1 & 1e-3 & 0.444 $\pm$ 0.042 & 0.736 $\pm$ 0.034 \\
21 & 256 & 128 & 1 & 1e-4 & \textbf{0.590 $\pm$ 0.044} & 0.486 $\pm$ 0.007 \\
22 & 256 & 128 & 3 & 1e-3 & 0.444 $\pm$ 0.021 & \underline{0.743 $\pm$ 0.045} \\
23 & 256 & 128 & 3 & 1e-4 & \underline{0.548 $\pm$ 0.033} & 0.493 $\pm$ 0.005 \\
24 & 256 & 128 & 5 & 1e-3 & 0.466 $\pm$ 0.007 & \textbf{0.902 $\pm$ 0.007} \\
\bottomrule
\end{tabular}
% }
\end{table}
\clearpage

\section{Results Experimental Setup}

% \begin{table}[ht]
%     \centering
%     \caption{Caption}
%     \begin{tabular}{|c|c|c|c|}
%     % \toprule
%     \hline
%        \textbf{Case}  & \textbf{s1-mel} & \textbf{s2-drum} & \textbf{s3-bass} \\
%        \textbf{}
%        \textbf{Prompt}  &  
%     \end{tabular}
%     \label{tab:placeholder}
% \end{table}

\lstdefinestyle{livecoding}{
    language=Ruby,
    basicstyle=\ttfamily\footnotesize,
    keywordstyle=\color{blue},
    commentstyle=\color{green!60!black},
    stringstyle=\color{red},
    numbers=none,
    breaklines=true,
    breakatwhitespace=true,
    tabsize=2,
    showspaces=false,
    showstringspaces=false,
    frame=none,
    backgroundcolor=\color{white}
}

\begin{table}[hb]
\caption{Code snippets and prompts used. For each snippet, GPT-5 generated 3 candidate and 10 candidate additions. The 10 candidates experienced a wider auditory variance, motivating the value of greater candidate generation and subsequent pruning with an embedding model. Our model successfully predicts the most distinct entries.}
\label{tab:experiment-setup}
\begin{tabular}{|p{1cm}|p{3.9cm}|p{4cm}|p{3.9cm}|}
\hline
 & \textbf{s1-mel} & \textbf{s2-drum} & \textbf{s3-bass} \\
\hline
\textbf{Code} & 
\begin{lstlisting}[style=livecoding]
use_bpm 100
# Drums
live_loop :drums do
  sample :bd_haus
  sleep 0.5
  sample :sn_dolf
  sleep 0.5
end
live_loop :hats do
  sleep 0.25
  sample :drum_cymbal
end
# Bassline
live_loop :bass do
  use_synth :fm
  play_pattern_timed [:e2, :g2, :a2, :g2], [0.5, 0.5, 0.5, 0.5], release: 0.25
end
\end{lstlisting} & 
\begin{lstlisting}[style=livecoding]
use_bpm 90
# Melody
live_loop :melody do
  use_synth :prophet
  play_pattern_timed [:c4, :e4, :g4, :e4], [0.5, 0.5, 0.5, 0.5], release: 0.3
end
# Bassline
live_loop :bass do
  use_synth :fm
  play_pattern [:g3, :g3, :d4, :g3], release: 0.5
end
\end{lstlisting} & 
\begin{lstlisting}[style=livecoding]
use_bpm 100
# Drums
live_loop :drums do
  sample :bd_haus   
  sleep 1
  sample :sn_dolf 
  sleep 1
end
# Melody
melody = [:e4, :g4, :a4, :b4, :a4, :g4, :e4, :d4].ring
live_loop :melody do
  use_synth :pulse
  play melody.tick, release: 0.3
  sleep 0.5
end
\end{lstlisting} \\
\hline

\textbf{Prompt} 
    & Propose $\{3,10\}$ melodies to accompany this code 
    & Propose $\{3,10\}$ drum patterns to accompany this code 
    & Propose $\{3,10\}$ bass lines to accompany this code \\
\hline

\end{tabular}
\end{table}

\end{document}